# Plasmon-assisted Förster resonance energy transfer at the single-molecule level in the moderate quenching regime

J. Bohlen[a,b,†], Á. Cuartero-González[c,†], E. Pibiri[a], D. Ruhlandt[d], A. I. Fernández-Domínguez[c], P. Tinnefeld[a,b,*], G. P. Acuna[a,e,*]

Metallic nanoparticles were shown to affect Förster energy transfer between fluorophore pairs. However, to date, the net plasmonic effect on FRET is still under dispute, with experiments showing efficiency enhancement and reduction. This controversy is due to the challenges involved in the precise positioning of FRET pairs in the near field of a metallic nanostructure, as well as in the accurate characterization of the plasmonic impact on the FRET mechanism. Here, we use the DNA origami technique to place a FRET pair 10 nm away from the surface of gold nanoparticles with sizes ranging from 5 to 20 nm. In this configuration, the fluorophores experience only moderate plasmonic quenching. We use the acceptor bleaching approach to extract the FRET rate constant and efficiency on immobilized single FRET pairs based solely on the donor lifetime. This technique does not require a posteriori correction factors neither a priori knowledge of the acceptor quantum yield, and importantly, it is performed in a single spectral channel. Our results allow us to conclude that, despite the plasmon-assisted Purcell enhancement experienced by donor and acceptor partners, the gold nanoparticles in our samples have a negligible effect on the FRET rate, which in turns yields a reduction of the transfer efficiency.

[a]Institute for Physical and Theoretical Chemistry – NanoBioScience and Braunschweig Integrated Centre of Systems Biology (BRICS), and Laboratory for Emerging Nanometrology (LENA), Braunschweig University of Technology, Braunschweig, Germany
[b]Faculty of Chemistry and Pharmacy, NanoBioScience, Ludwig-Maximilians-Universität München, München, Germany
[c]Departamento de Física Teórica de la Materia Condensada and Condensed Matter Physics Center (IFIMAC), Universidad Autónoma de Madrid, E-28049 Madrid, Spain
[d]Third Institute of Physics – Biophysics, Georg-August-Universität Göttingen, Göttingen, Germany.
[e]Department of Physics, University of Fribourg, Chemin du Musée 3, Fribourg CH-1700, Switzerland.
†Contributed equally
*Corresponding author: philip.tinnefeld@cup.lmu.de, guillermo.acuna@unifr.ch

## 1. Introduction

Surface plasmons supported by metal nanostructures can affect the photophysical properties of fluorophores in multiple ways.[1,2] First, they can alter the excitation rate by changing the intensity of the incident electric field at the fluorophore's position.[3] Second, they can modify the radiative and non-radiative decay rates of molecules through the photonic local density of states, thus affecting their overall quantum efficiency and fluorescence lifetime.[4] Finally, surface plasmons can also shape the fluorophore emission pattern into the far-field.[5,6] Over the last decades, these abilities of metal nanoparticles (NPs) were exploited for the development of optical antennas,[7–9] which have enabled nanophotonic applications ranging from fluorescence enhancement[3,10,11] or photostability[12,13] increment to the detection of single molecules at elevated concentrations[14–17] or the sequencing of DNA in real time.[18]

Förster (or fluorescence) resonance energy transfer (FRET) is the non-radiative dipole-dipole energy exchange between two (donor and acceptor) fluorophores. The extreme sensitivity of this mechanism to the inter-molecular distances (in the few nanometer range) is currently being exploited in a wide range of biophysical and cell biological[19,20] tools, which make it possible to monitor the change in conformation and structure of biological complexes. Moreover, FRET also plays a fundamental role in light harvesting processes[21,22] in plants and photosynthetic bacteria. Apart from its fundamental interest, a profound understanding of FRET and its photonic implications is expected to be instrumental for the development of highly efficient organic photovoltaic devices.[23,24]

Recent theoretical[25,26] and experimental[27] studies indicate that metal structures can alter the energy transfer between donor-acceptor fluorophore pairs, enlarging the energy-transfer distance,[28] and improving fluorescence image resolution.[29] However, the net effect of surface plasmons on FRET remains controversial.[30] Contradictory phenomena have been reported ranging from FRET efficiency reduction[31–33] and enhancement,[27,28] together with a linear and non-linear dependence of the FRET rate on the photonic local density of states.[30,34,35] This lack of conclusive results and overall agreement can be attributed mainly to two factors. First, it is challenging to position a FRET pair in the near field of a metallic nanostructure with nanometer precision. Second, it is also extremely demanding to isolate the effect of the surface plasmons supported by metal NPs on FRET. Indeed, most studies were performed at the ensemble level based on an analysis of both the donor and acceptor intensities. Thus, the FRET rate and efficiency were extracted from averaged populations and not for each single fluorophore pair. Furthermore, these approaches required correction factors and previous knowledge of the NPs effect on the donor and acceptor quantum yield. Note that the fluorophore-NP interaction is characterized by a strong spectral dispersion, as extensively reported in the literature for the fluorescence intensity[11,36] and lifetime[37] enhancement and quenching. This is due, among other factors, to its dependence on the NP size and shape and on the relative orientation of the fluorophore and its distance to the NP. Therefore, conclusive results can only be drawn if FRET is studied at the single NP-fluorophore pair level. These limitations call for a thorough alternative strategy to settle the plasmon-assisted FRET controversy.

In this work, we use the DNA origami technique to position single FRET pairs 10 nm away from single Au NPs of different sizes. These NPs exhibit an extinction cross section that overlaps with the absorption and emission spectral ranges of the donor (strongly) and acceptor (moderately) fluorophores. Experimental reports indicate that 10 nm is the distance where fluorescence quenching of molecules by metal particles is roughly 50 % and therefore this is a very relevant and sensitive distance range.[38–40] We determine, at the single molecule level and on immobilized fluorophore pairs, how the NP size affects the FRET rate and efficiency. Our results, obtained following the so-called "acceptor bleaching" approach, allow us to conclude that for sizes between 5 and 20 nm, despite the significant Purcell enhancement experienced by donor and acceptor, there is no significant change in the FRET rate between them. Therefore, the FRET efficiency is reduced due to the increment of the total decay rates of fluorophores in the vicinity of Au NPs. Our findings are supported by electromagnetic calculations implementing a semi-classical model for FRET, parameterized according to the experimental samples and yielding excellent agreement with measured results.

## 2. Sample preparation and FRET characterization

The study of plasmon-assisted FRET proves to be significantly challenging. The first difficulty comprises sample fabrication. Although the first pioneering experiments were performed on an undetermined number of FRET pairs in the near field of NP dimers,[41] a detailed understanding of the plasmon-assisted FRET effect demands the fabrication of single donor-acceptor pairs with a controlled intermolecular distance, as well as their precise positioning nearby a metal nanostructure. DNA as a scaffold has been extensively employed to self-assemble FRET pairs with nanometer precision,[42] through the hybridization of two complimentary single DNA strands labeled with a donor and acceptor fluorophore respectively. In fact, Wenger and coworkers have exploited this approach to reveal how zero-mode waveguides[32] (also termed nano-apertures) and dimer optical antennas fabricated within nano-apertures[43] modify the FRET of diffusing donor-acceptor pairs based on double-stranded DNA sequences in solution. This approach was also employed to fix the relative orientation between donor and acceptor.[27,31,33] These pioneering works were only able to account for the spatially averaged effect of the metallic structures on FRET because the donor-acceptor pair was allowed to freely diffuse within the nano-apertures. Recently, double-stranded DNA was also employed to place a FRET pair at the hotspot of an optical antenna based on one and two Au NPs.[31,33] The introduction of the DNA origami technique[44] enables the self-assembly of complex hybrid structures, in three dimensions, where different species such as dye molecules, quantum dots, and metal NPs can be positioned with nanometric precision and stoichiometric control.[45] Thus, it has been exploited for nanophotonic applications in recent years[46–48] including the study of FRET in the vicinity of Au NPs.[34]

The second obstacle for FRET assessment originates from the far-field measurement method itself, and the indirect extraction of the transfer rate and efficiency near metal NPs. Note that the FRET efficiency $E$ is defined as[49]

$$E = 1 - I_{DA}/I_D = \frac{I_{AD}/\phi_A}{I_{AD}/\phi_A + I_{DA}/\phi_D} \qquad (1)$$

where $I_D$ and $I_{DA}$ are the fluorescence intensities of the donor fluorophore in the absence and presence of the acceptor respectively, $I_{AD}$ the acceptor's fluorescence intensity upon donor excitation and $\phi_A$ ($\phi_D$) the quantum yield of the acceptor (donor). The central and right hand side of equation (1) enable the calculation of $E$ with different experimental approaches. In the central expression, only the fluorescence intensity of the donor needs to be measured in a single spectral channel. However, it is necessary to determine it in the presence and absence of the acceptor. In experiments with single immobilized molecules, this is typically achieved by waiting until the acceptor bleaches (acceptor bleaching approach). In contrast, the expression on the right side requires the measurement of the fluorescence signal of both donor and acceptor, and therefore in two different spectral channels.

Similarly to Equation (1), the FRET rate constant $k_{ET}$ can be estimated from the donor's fluorescence lifetime in the presence $\tau_{DA}$ and absence of the acceptor $\tau_D$ as

$$k_{ET} = \frac{1}{\tau_{DA}} - \frac{1}{\tau_D} \qquad (2)$$

It is worth noticing that for a particular FRET pair, and under the same excitation and detection conditions, $I_{DA}/I_D = \tau_{DA}/\tau_D$ and therefore the FRET efficiency can also be determined based on fluorescence lifetime measurements as[39]

$$E = 1 - \frac{\tau_{DA}}{\tau_D} \qquad (3)$$

Note that lifetime measurements are typically more reliable than intensity measurements since they do not depend on the analyte concentration and instrument alignment, neither they are sensitive to saturation effects.

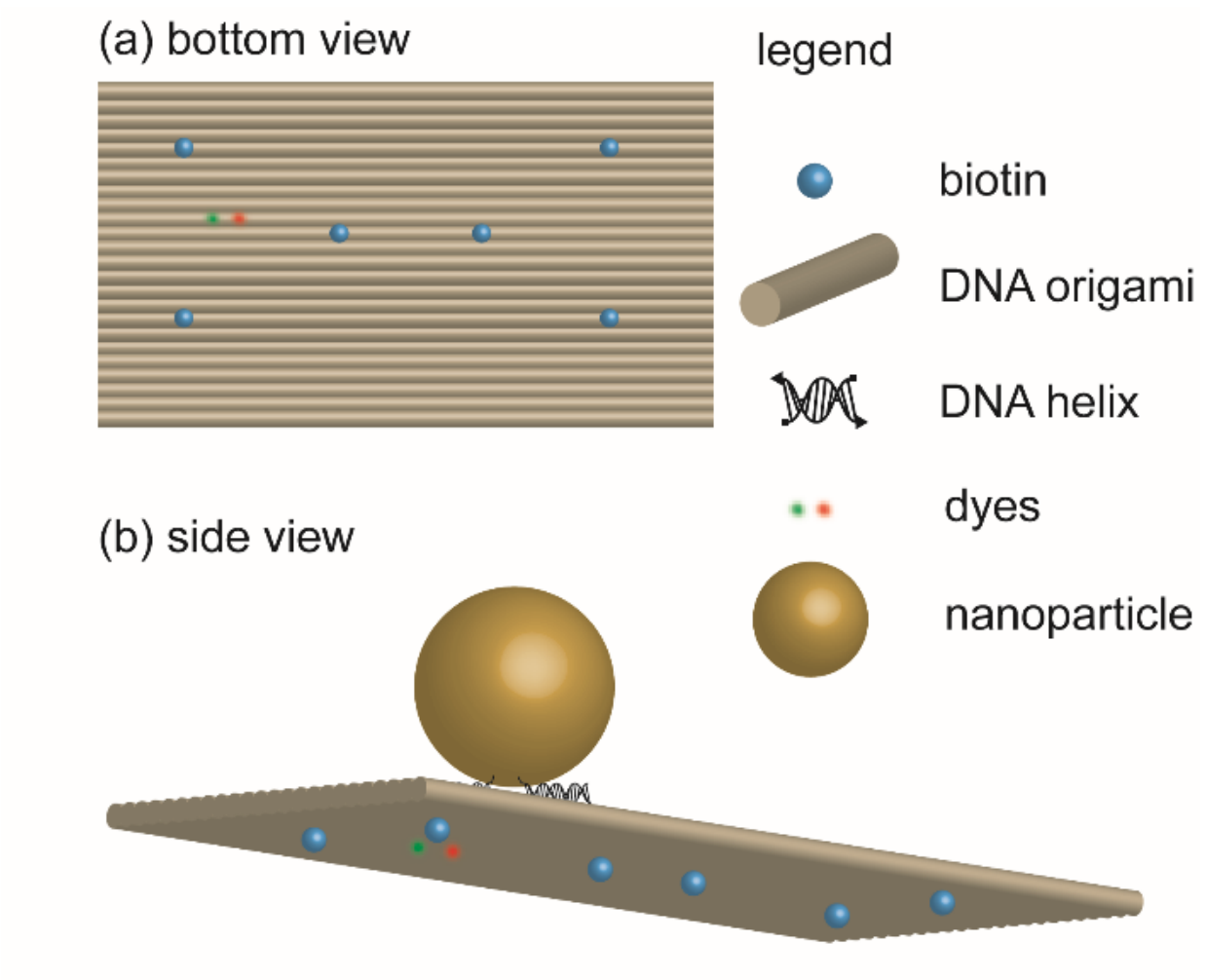

Figure 1: Sketch of the rectangular DNA origami structure. (a) The bottom view shows the FRET pair of dyes and the six biotins for the surface immobilization. (b) Side view depicting the capturing strands employed for the incorporation of a single metal NP.

As discussed above, to date, the acceptor bleaching approach has not been employed to study plasmon-assisted FRET at the single-molecule level. Instead, most experiments were performed at the ensemble level and on freely diffusing FRET pairs in solution. Ensemble measurements have the inherent disadvantage that only averages over populations can be studied. This is particularly relevant for FRET measurements in which factors like the presence of impurities (including for instance colloidal NP aggregates), or defective plasmonic NP-FRET-pair structures (such as, for example, those where only donor or acceptor are present, where the acceptor is bleached, or the NP is missing) can severely affect the overall results. Furthermore, within ensemble measurements of freely diffusing FRET pairs, the photophysics of a single donor in the presence and absence of its acceptor counterpart cannot be monitored, and therefore the right part of Eq. (1) has to be employed. For plasmon-assisted FRET measurements, this approach has the additional shortcoming that the plasmonic nanoparticles affect the donor and acceptor quantum yields $\phi_D$ and $\phi_A$, respectively, thus greatly complicating the reliable determination of $E$. Finally, a few studies were performed on immobilized samples, but the FRET efficiency was obtained from the intensities of the donor and acceptor channels.[33] In another experiment, FRET rate constants and efficiencies were extracted by comparing the average donor's lifetime on two samples with and without acceptor[30] at the ensemble level.

In order to overcome the aforementioned limitations, we here employ the DNA origami technique to position both the metal NP and the FRET pair and perform single-molecule fluorescence measurements on the resulting surface-immobilized samples. Fig. 1 includes a sketch of these samples, based on a rectangular DNA origami structure with dimensions of 70 nm x 85 nm (the thickness of a DNA double-helix is approximately 2 nm). The FRET pair consists of ATTO532 (donor) and ATTO647N (acceptor) molecules.[38] It is attached to the DNA origami structure through internal labelling on the same double helix,[50] see Fig. 1(a), resulting in a gap of approximately 3.4 nm between the fluorophores. Six biotin-functionalized oligonucleotides are used to immobilize the DNA origami structure on a glass coverslip, which is functionalized with BSA-biotin and neutrAvidin. Following surface immobilization, a single metal NP is bound at a predefined position on the upper side of the origami through DNA hybridization,[41] see Fig. 1(b). We employed 5, 10, 15 and 20 nm Au NPs. The FRET pair is located at the bottom side to avoid physical contact of dyes and nanoparticle. The distance between the NP surface and the FRET pair is approx. 10 nm (based on geometric calculations assuming the length of each nucleotide to be 0.34 nm). For these NPs´sizes and distances to the FRET pair, the fluorescence lifetime reduction can be accurately determined. Further details on sample fabrication can be found in the Methods and Materials section, whereas a table containing the distances between NPs and fluorophores can be found in the Supporting Information (SI).

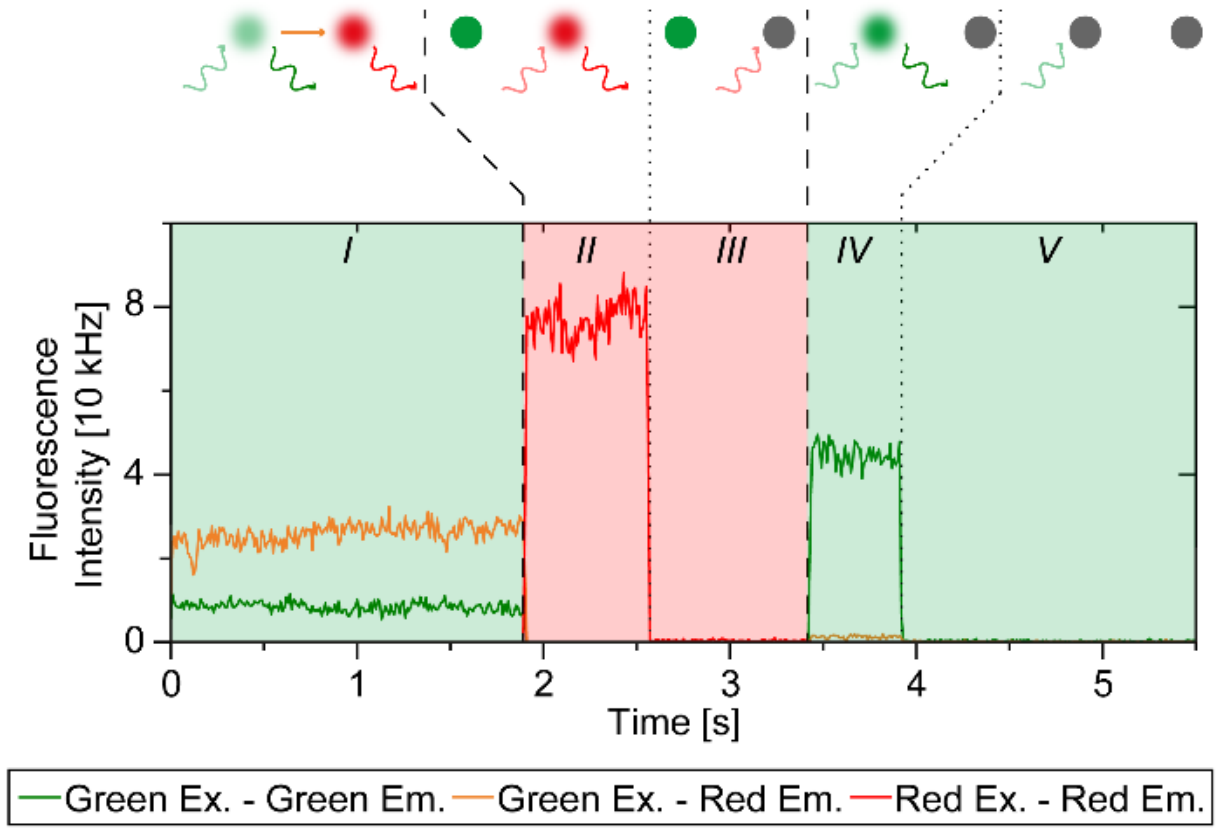

Figure 2: Example of a fluorescence transient obtained through laser alternation for single-molecule FRET determination using the "acceptor bleaching" approach. In *I*, only the green laser is on to monitor $\tau_{DA}$, $I_{DA}$ (green excitation –green detection) and $I_{AD}$ (green excitation –red detection). In *II* and *III*, only the red laser is switched on to measure $I_A$ and $\tau_A$ (red excitation –red detection) until the acceptor bleaches (*III*). In *IV* and *V*, only the green laser is exciting to determine $\tau_D$ and $I_D$ until the donor bleaches (*V*).

Samples were scanned with a home-built confocal fluorescence microscope in order to locate the immobilized structures. For each FRET pair, fluorescence transients were recorded. In order to maximize the amount of information that can be extracted from fluorescence transients, we manually alternated between donor and acceptor excitation. This procedure is illustrated in Fig. 2. Initially, the donor is excited (*I*, donor excitation at 532 nm), this allows us to extract $I_{DA}$ (green transient, donor intensity upon donor excitation), $I_{AD}$ (orange transient, acceptor intensity upon donor excitation) and the fluorescence lifetime $\tau_{DA}$. Afterwards, the sample is excited in the red spectral range (*II*, acceptor excitation, 640 nm) to determine $I_A$ (red transient, acceptor intensity upon acceptor excitation) and its corresponding fluorescence lifetime $\tau_A$, until the acceptor is bleached in *III*. Finally, the sample is excited again in the green spectral range, *IV*, now to record $\tau_D$ and $I_D$ until the donor bleaches (*V*). Importantly, this technique enables the determination of the background signal in each channel and the verification (through the single bleaching steps) that the fluorescence measured arises from single FRET pairs. The presence of single Au NPs can be independently inferred by the reduction of $\tau_D$ and $\tau_A$ as Au NPs quench both the acceptor and the donor.[51] This procedure was repeated for DNA origami structures with no NPs for referencing. In order to rationalize our experimental results, we perform numerical electromagnetic simulations modelling our system. We use measured values for all geometric parameters (NP radii, DNA origami thickness, and dye-NP and intermolecular distances). Au permittivity is taken from experimental data[52] and the refractive index of DNA origami structure is set to 2.1.[53] We carry out three different numerical studies. In the first two, only one molecule (donor or acceptor) is included as a point-dipole-like electromagnetic source. By averaging over three perpendicular dye orientations, we compute the total Purcell spectrum for all the experimental geometries. Performing the spectral average within the dye emission window and taking into account its intrinsic quantum yield $\phi_{D,A}$, we obtain the fluorescence lifetimes $\tau_D$ and $\tau_A$ and investigate their sensitivity to the Au NP size (see SI). In the third study, the donor is treated again as a dipole source, but the acceptor is modelled as a dielectric sphere whose randomly oriented polarizability matches the one orresponding to a quantum two-level system[44] (see the Methods and Materials for further details). These simulations yield the donor Purcell factor in the presence of the acceptor, from which we determine $\tau_{DA}$. Combining these results with those in the absence of the acceptor, we obtain the FRET efficiency *E* from Equation (3). In addition, we also calculate the FRET rate constant $k_{ET}$ using Equation (2) or directly by computing the spatial average of the electric field intensity within the dielectric sphere modelling the acceptor molecule,[54,55] $k_{ET} \propto V^{-1} \int |E_{DA}|^2 dV$.

## 3. Results and discussion

Figure 3 (a), (d) and (g) shows the sample-averaged fluorescence intensities $I_A$, $I_D$ and $I_{DA}$ for different NP diameters. All values were normalized to the intensity obtained without NPs (the measured distributions can be found in the SI).

Figure 3: Summarized results of the measurements with standard deviation and simulation: The normalized averaged fluorescence intensity against the nanoparticle diameter for the donor in presence ($I_{DA}$; (a)) and absence of the acceptor ($I_D$; (d)) and the acceptor only ($I_A$; (g)). The fluorescence lifetime measurements are shown in hollows symbols with error bars compared to the simulated results (filled symbols and indicated by the index *sim*) for the donor with acceptor ($\tau_{DA}$, $\tau_{DA,\,sim}$; (b)), after photobleaching of the acceptor ($\tau_D$, $\tau_{D,\,sim}$; (d)) and acceptor only ($\tau_A$, $\tau_{A,\,sim}$; (h)). The calculated and simulated FRET efficiency ($E$, $E_{sim}$) and FRET rate ($k_{ET}$, $k_{ET,\,sim}$) are diagrammed in (c) and (f). The difficult differentiation between simulated and experimental results shows a good agreement between both data sets.

It is worth mentioning that the distance between the NP surface and the donor (acceptor) decreases slightly with the NP size, from 10.48 (11.39) nm for 5 nm NPs to 8.76 (9.41) nm for the 20 nm NPs (all distances can be found in the SI). As previously observed, for fluorophores located under the "polar" plane of the NP as defined by the incident light polarization, the overall reduction of the quantum yield due to an increment of the non-radiative rate prevails over the increment of the excitation rate.[4,40] As a result, a reduction of the fluorescence intensity is measured. This effect is stronger in $I_D$ than in $I_A$ due to the spectral overlap between the donor emission and the Au NPs resonance[11] in the green spectral range. In the case of $I_{DA}$, FRET to the acceptor in close proximity prevails, and the effect of the plasmonic NP on the intensity at the donor channel is significantly less pronounced. Figure 3 (b), (e) and (h) plot measured (empty dots) and simulated (solid dots) fluorescence lifetimes $\tau_{DA}$, $\tau_D$ and $\tau_A$. Remarkably, both are in very good agreement with theoretical predictions lying within the experimental error bars in all cases. The data sets are normalized to the samples without NPs and are also presented in absolute scale (see right axis), revealing up to a two-fold (four-fold) total Purcell enhancement for the acceptor (donor) molecules. These results show a similar trend as the intensities in Figure 3 ((a), (d) and (g)), which is in accordance with previous reports.[40] Note again the quenching visible in $\tau_D$, which takes place in the green region of the electromagnetic spectrum. The presence of the metal NPs accelerates the decay of both dyes, with a stronger effect on the non-radiative channel. Therefore, the overall effect on the fluorescence lifetime is comparable to the one on the quantum yield. As the increment in the excitation rate (electric field enhancement at the dyes position) is negligible, similar reductions of the fluorescence lifetime and of the intensity are observed as previously reported. As in Fig. 3 (a), (d) and (g), the comparison between $\tau_{DA}$ and $\tau_D$ in Fig. 3 (b), (e) and (h) demonstrates that the presence of the acceptor diminishes the effect of the Au NPs in the donor fluorescence characteristics.

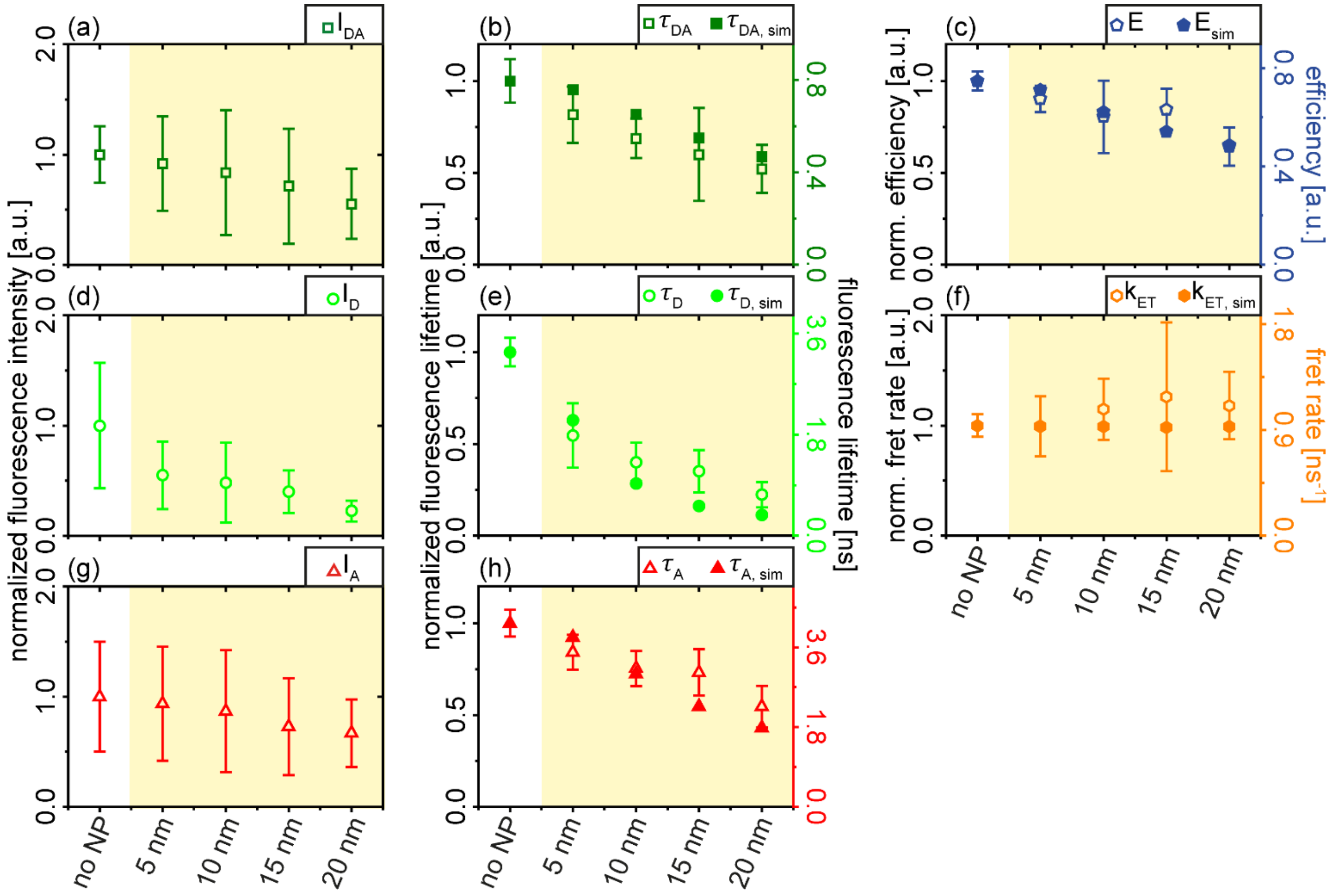

Introducing the measured donor lifetimes $\tau_{DA}$ and $\tau_D$ into Equations (2) and (3), we can extract the FRET rate constant $k_{ET}$ and FRET efficiency $E$ for each single donor-acceptor pair in the presence of Au NPs. The experimental results obtained this way and normalized to the results of samples without NPs are shown as empty dots in Figure 3 (c) and (f). Electromagnetic calculations for these two magnitudes are plotted in solid dots. Similar to the experiments, the numerical FRET efficiencies are computed by evaluating Equation (3) using the theoretical predictions for $\tau_{DA}$ and $\tau_D$. On the contrary, as discussed above, the FRET rates in Figure 3 (c) and (f) are calculated directly from simulations through the spatial averaging of the electric field intensity within the acceptor volume. The agreement between this direct estimation for $k_{ET}$ and an indirect one, consisting in the evaluation of Equation (2) through numerical data, is shown in the SI. Both numerical and experimental results indicate that the presence of the metal nanostructure does not have a significant impact on the FRET rate constant. We can observe that Au NPs decrease the FRET efficiency, being the reduction in $E$ of 25% for the largest structure (20 nm diameter). Note that, according to Equations (2) and (3), $E = k_{ET}\,\tau_{DA}$, which reveals that the decrease of the FRET efficiency in Figure 3 (c) is a direct consequence of the reduction of the donor lifetime in presence of the metal NP and acceptor molecule in Figure 3 (b). Importantly, the simple expression above also clarifies why $E$ is not significantly modified due to the metal NP, despite the Purcell lifetime reduction experienced by the donor molecule. It shows that $\tau_{DA}$ is the time scale that sets the transfer efficiency, and it is less sensitive to the plasmon field than $\tau_A$.

## 4. Conclusion

In summary, we have exploited the DNA origami technique to self-assemble structures where a single Au NP and a fluorescent donor-acceptor pair were positioned with stoichiometric control and nanometer precision. These structures were used to analyze the effect on the FRET induced by Au NPs of different diameters (ranging from 5 to 20 nm) placed 10 nm away from the fluorescent pair, which is separated by 3.4 nm. Our measurements were performed at the single-molecule level on surface immobilized structures using the “acceptor bleaching” technique. This approach enabled the reliable determination of the plasmon-assisted FRET rate and efficiency based solely on the measurement of the donor`s fluorescence lifetime in the presence/absence of the acceptor. The experimental results are supported by electromagnetic calculations implementing a semiclassical model for FRET. Our findings contradict previous works using colloidal NPs and DNA, in which an enhancement of the FRET rate with the LDOS was.[31,34] The presented measurements, performed at the single molecule level following the “acceptor bleaching” technique, reveal that, despite the significant plasmon-assisted fluorescence lifetime reduction and quenching experienced by both donor and acceptor molecules, the Au NPs have a minor effect on the FRET rate in our experimental samples. In contrast, the FRET efficiency decreases with increasing NP size through the fluorescence lifetime reduction undergone by the donor fluorophore in presence of the NP and its acceptor counterpart.

## 5. Material and Methods

If no other company is mentioned all chemicals were ordered by Sigma Aldrich.

#### A. Preparation of DNA origami structures

The rectangular DNA origami structures were produced by adding the unmodified, modified staples (including the oligonucleotides with Biotin, Atto647N, Atto532 and capturing strands for the nanoparticle), the folding buffer (final concentration: 1 x TAE, 12 mM $MgCl_2$) and the scaffold p7249 (final concentration: 27.2 nM). The modified and unmodified staples had a tenfold concentration compared to the scaffold. To fold the DNA origami structures the following program was used: Heating up to 70 °C for 5 min and then cooling down with a temperature gradient of - 1 °C/min to a final temperature of 24 °C.
Gel purification was used to separate the oligonucleotides from the DNA origami structures. The gel consists of 1.5 % vol agarose (Biozym LE Agarose) and 50 mL TAE (1 x TAE with 12 mM $MgCl_2$ x 6 $H_2O$). Also 2 µL peqGreen (VWR) were added to the Gel and 1 x BlueJuice (Thermo Fisher Scientific) as a loading buffer for the sample. As the gel buffer 1 x TAE with 12 mM $MgCl_2$ is used. The total run time for the cooled gel was 2 hours with a voltage of 80 V. An example of a gel is shown in figure S4 in the supplementary information.
The correct folding of the DNA origami structures was characterized with atomic force microscopy (AFM, Nanowizard 3 ultra, JPK Instruments) in solution. On a freshly cleaved mica surface (Qualty V1, Plano GmbH) 10 µL of a 10 mM $NiCl_2$ x 6 $H_2O$ solution were incubated for 5 min. After three times washing with 300 µL miliQ-water (Merck Milli-Q) and drying with compressed air, 10 µL 1 nM DNA origami structure solution (diluted in AFM buffer (40 mM TRIS, 2 mM EDTA disodium salt dihydrate and 12.5 mM $Mg(OAc)_2$ x 6 $H_2O$) were added and incubated for 5 min. Afterwards 300 µL AFM buffer were added after purging three times with 300 µL AFM buffer. The solution measurements were performed with cantilevers USC-F0.3-k0.3-10 from Nano World. AFM images of the DNA origami structures with and Au NPs are included in figures S4 and S5 respectively.

#### B. Functionalization of nanoparticles

Au NPs were ordered from BBI solutions and functionalized with 25T single-stranded DNA oligonucleotides (Ella Biotech GmbH) labelled with a thiol group at 3´-end. After cleaning the coated stir bars, glass and snap on lid with ultra-pure water (Merck Milli-Q), a 2 mL NP solution was added. To the stirred solution (550 rpm), 20 µL Tween20 (10%, Polysorbate20, Alfa Aesar), 20 µL of a potassium phosphate buffer (4:5 mixture of 1 M monobasic (P8709) and dibasic potassium phosphate (P8584)) and an excess of 50 nM oligo (for the volume see supplementary information) were added. After heating the solution for one hour at 40 °C, the solution was salted every 3 minutes with a PBS solution containing 3.3 M NaCl to a final concentration of 750 mM NaCl. For the followed salting steps see supplementary material.

**C. Sample preparation**

Lab-Tek chambers (Thermo Scientific) were incubated for 2 min with 200 µL 0.1 M hydrofluoric acid (AppliChem), washed three times with 300 µL NP buffer (1 x TAE, 12.5 mM $MgCl_2$, 300 mM NaCl) and incubated again with 200 µL 0.1 M hydrofluoric acid. The hydrofluoric acid provides a clean surface. After cleaning three times with 300 µL NP buffer 100 µL BSA-Biotin (1 mg/mL) is added and incubated overnight at 4 °C. The BSA-Biotin passivates the surface against unspecific binding. The next day the surface is washed three times with 300 µL NP buffer. Afterwards 100 µL neutrAvidin (1 mg/mL) is added and incubated for 10 min, the surface is washed three times with 300 µL NP buffer. 200 µL DNA origami structures solution (~80 pM) is added, the surface density is monitored with the confocal setup. After cleaning the surface three times with 300 µL NP buffer, 200 µL SuperBlock (PBS) blocking buffer (Thermo Scientific) is added for 10 min to achieve additional surface passivation. Following the purging of the surface with three times 300 µL NP buffer the nanoparticle solution is added and incubated for 48 h at 4 °C. The NP absorption was set to 0.05 and monitored at a UV-vis spectrometer (Nanodrop 2000, Thermo Scientific). Finally, after washing three times with 300 µL NP buffer to get rid of the nanoparticles in solution, a trolox/trolox quinone solution is added to increase photostability.[56]

**D. Imaging**

Single molecule fluorescence measurements were performed at a custom-build confocal setup based on an Olympus IX-71 inverted microscope. As power sources a 637 nm (LDH-D-C-640, Picoquant) and a 532 nm (LDH-P-FA530B) pulsed laser are used with an intensity for the FRET samples of 9 µW and 2 µW respectively. Both lasers beams were modified by an AOTF filter (AOTFnc-VIS, AA optoelectronic), cleaned up and expanded by an optical fiber, before entering a λ/2 (LPVISE100-A, Thorlabs) and a λ/4 (AQWP05M-600, Thorlabs) plate to achieved circularly polarized light. A dichroic mirror (Dualband z532/633, AHF) was employed to direct the beam to an oil-immersion objective (UPLSA-PO100XO, NA 1.40, Olympus). A piezo stage (P-517.3CL, Physik Instrumente GmbH & co. KG) scans the sample by moving the Lab-Tek over the objective. In this scan every molecule can be selected to perform a time-resolved analysis. The emitted fluorescence is collected by the objective and separated from the excitation light through the dichroic mirror. To minimize the detection volume the beam is focused through a pinhole (Linos 50 µm). The fluorescence light is divided by a dichroic mirror (640DCXR, AHF) and the red and green emission is purified with different filter, Bandpass ET 700/75m, AHF; RazorEdge LP 647, Semrock (red) and Brightline HC582/75, AHF; RazorEdge LP 532, Semrock (green). Both signals are detected at different Diodes (τ-SPAD-100, Picoquant) and the time-resolved analysis is done by a single-photon counting card (SPC-830, Becker&Hickl). The raw data analysis is performed by a home written LabView software (National instruments).

**E. Theoretical model and calculations**

In order to verify the experimental results, we have performed numerical simulations using the finite-element solver of Maxwell´s Equations in the commercial software COMSOL MULTIPHYSICS™. First, conventional Purcell factor, $Pf$, calculations for the donor and acceptor molecules were carried out for all the relevant orientations. In these simulations, the power radiated through a small box including only the dipole source was computed within a frequency window matching the experimental emission spectra. The dye lifetime $\tau_i$ with *i = D, A* was then extracted through spectral averaging, and taking into account the inherent quantum yield $\phi_i$

$$\tau_i = \frac{\tau_i^{(0)}}{\phi_i \, Pf_i - (1 - \phi_i)} \qquad (4)$$

where $\tau_i^{(0)}$ is the lifetime in vacuum (absence of the Au NP).

Simulations describing the emission of the donor in the presence of the acceptor were also performed. In these calculations, a semiclassical model for FRET was implemented, in which the donor is treated as dipole-like electromagnetic source and the acceptor is effectively described as an absorbing dielectric sphere. This is similar to a model recently proposed in the context of plasmon-assisted exciton transport[54] and strong coupling.[55] The randomly oriented polarizability of this sphere is set to match the polarizability of a quantum two-level system. The resulting effective dielectric function has the form

$$\varepsilon_{A,eff}(\omega) = \frac{1 - 2\,\eta_A(\omega)}{1 + \eta_A(\omega)} \qquad (5)$$

with

$$\eta_A(\omega) = \frac{\mu_A^2\ \omega_A}{3\ \varepsilon_0\ V\ \hbar\ \omega\left(\omega - \left(\omega - \frac{i\,\gamma_A}{2}\right)\right)} \qquad (6)$$

where $\varepsilon_0$ is the vacuum permittivity and $V$ the sphere volume. Three parameters, set in accordance with experiments, were required to describe the acceptor molecules: dipole moment ($\mu_A = 14.5$ D), natural frequency ($\omega_A = 1.9$ eV), and linewidth ($\gamma_A = 0.1$ eV). The convergence of results against $V$ was checked (the radius of the sphere was finally set to 0.25 nm). Note that this simplified model does not account for the stoke shift of ATTO647N, and that the absorption spectrum resulting is purely Lorentzian while the actual profile presents a well-defined vibronic sideband.

## Conflicts of interest

There are no conflicts to declare.

## Funding

Deutsche Forschungsgesellschaft (DFG) (AC 279/2-1, TI 329/9-1); Spanish MINECO (FIS2015-64951-R and MDM-2014-0377-16-4), European 7[th] Framework programme (CIG-630996), European Research Council (ERC) (SiMBA, EU 26116), ChipScope and excellent cluster of Ludwig-Maximilians-Universität München Center for Integrated Protein Science Munich (CIPSM) and Nanosystems Iniative Munich (NIM). This work was supported by the Swiss National Science Foundation through the National Center of Competence in Research Bio-Inspired Materials.

## Acknowledgements

We thank Kristina Hübner and Johannes Feist for fruitful discussions, and Tim Schröder for the setup support.

## Notes and references

# SI: Plasmon-assisted Förster resonance energy transfer at the single-molecule level in the moderate quenching regime

J. Bohlen[a,b,†], Á. Cuartero-González[c,†], E. Pibiri[a], D. Ruhlandt[d], A. I. Fernández-Domínguez[c], P. Tinnefeld[a,b,*], G. P. Acuna[a,e,*]

## 1. Spectra

An overview of all spectra, including scattering and absorption of the monomer nanoparticle and absorption and emission of the FRET pair are shown in figure S1. The data for the nanoparticles are computed with the Mie Theory Calculator from Nanocomposix and the dye spectra are from the Atto tec website.

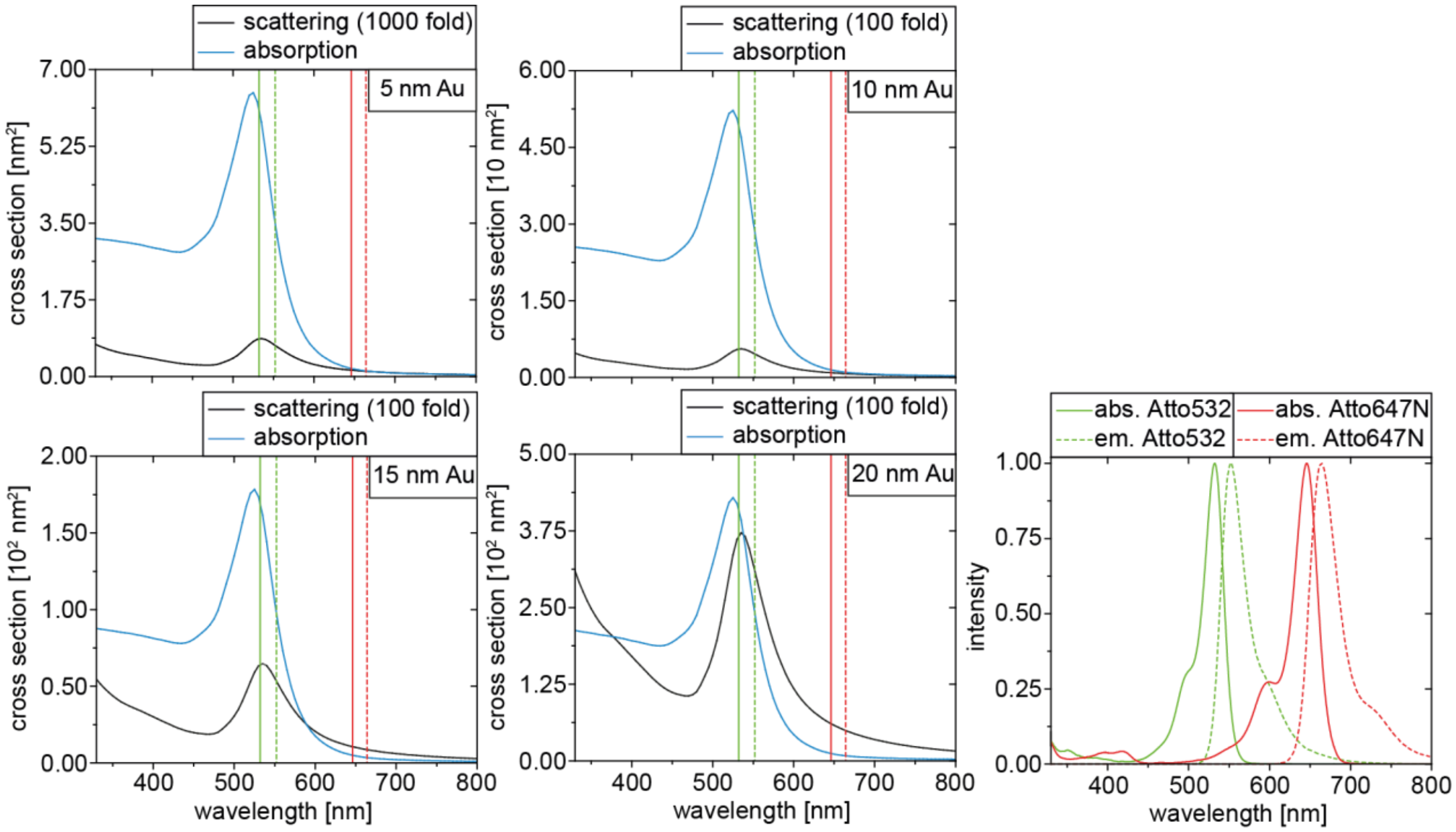

Figure S1: Scattering (black) and absorption spectra (blue) of the employed nanoparticles with the absorption (continuous) and emission maxima (dashed) of the Atto532 (green) and Atto647N (red). In addition, the whole spectra of the FRET pair is diagrammed.

## 2. Raw Data

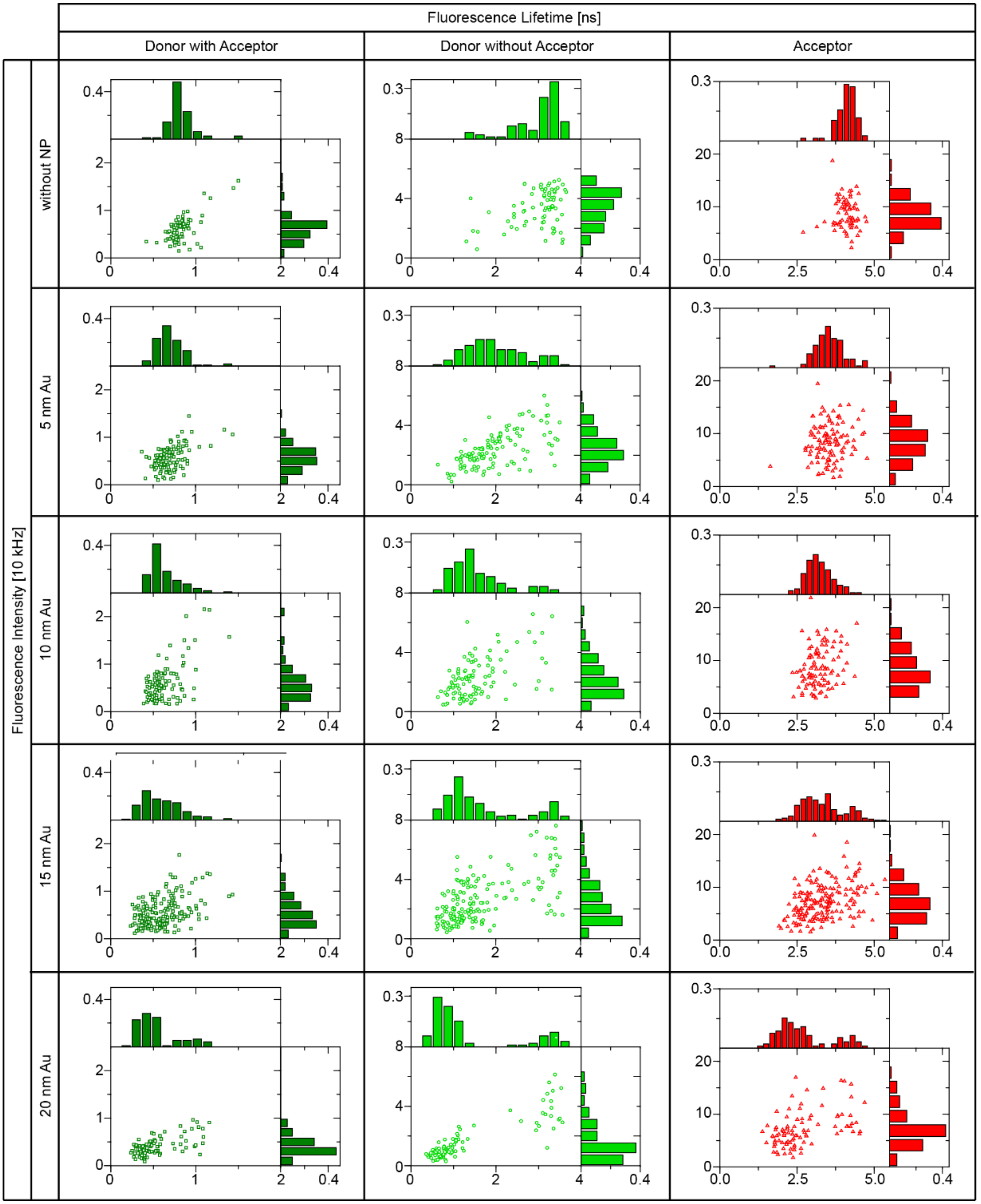

Figure S2: Raw data of the fluorescence lifetime and intensity of all three channels (donor in the presence of the acceptor and after photobleaching of the acceptor and acceptor only) from the measured with and without nanoparticle.

## 3. Distance calculation between dyes and nanoparticle surface

For the distance between dyes and nanoparticle a, the centroid (S) of the fictive triangle between all possible capturing strands ($P_1$, $P_2$, $P_3$) has to be calculated (see Figure S3).

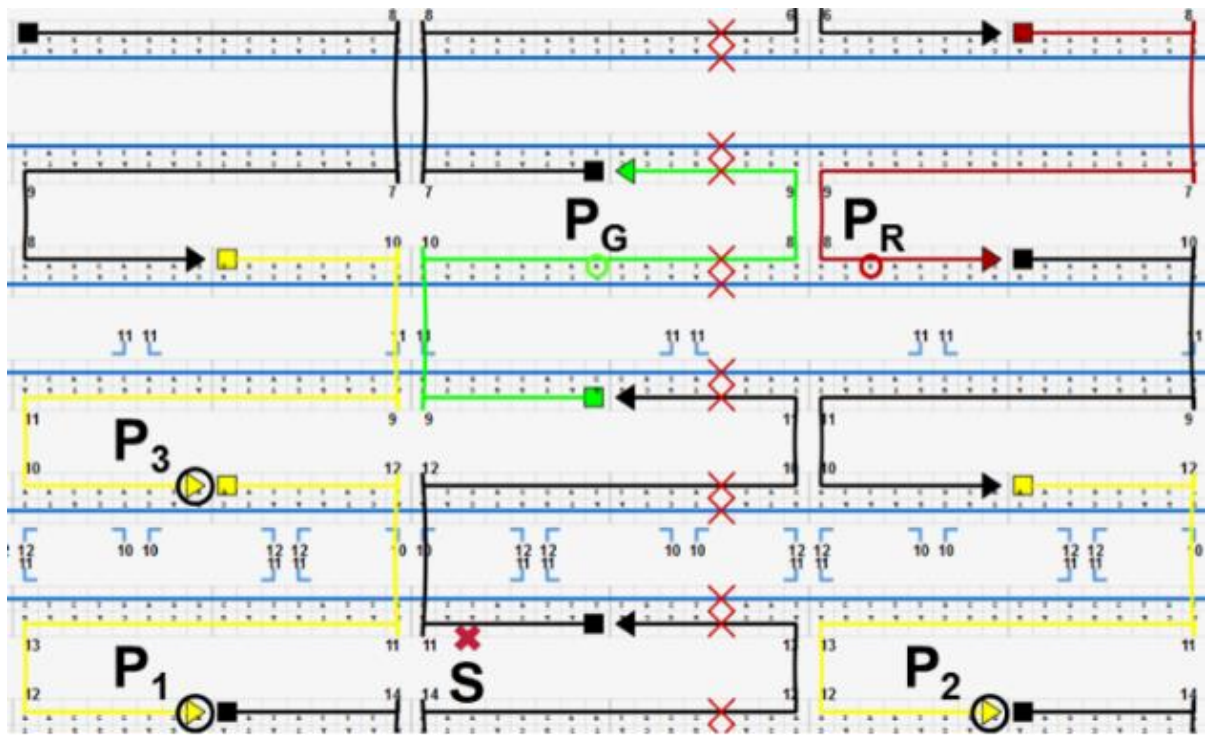

Figure S3: Section from the caDNano images with positions of Atto647N (PR), Atto532 (PG), all capturing strands (P1, P2, P3) and centroid of the capturing strands (S).

With the equations (S1) and coordinates (see table S 1) the centroid S ($x_S$, $y_S$) can be calculated.

$$x_S = \frac{x_{P_1} + x_{P_2} + x_{P_3}}{3};\ y_S = \frac{y_{P_1} + y_{P_2} + y_{P_3}}{3} \qquad (1)$$

Table S1: coordinates of Atto647N ($P_R$), Atto532 ($P_G$), all capturing strands ($P_1$, $P_2$, $P_3$) and centroid of the capturing strands (S) (the n in the index Indicates a Position, e.g. $x_{P_1}$ stands for the x coordinate of $P_1$.

| | $P_1$ | $P_2$ | $P_3$ | S | $P_R$ | $P_G$ |
|---|---|---|---|---|---|---|
| Helix ($x_n$) | 13 | 13 | 11 | 12.3 | 9 | 9 |
| Base ($y_n$) | 63 | 94 | 63 | 73.3 | 89 | 79 |

Distances between S and $P_R$ or $P_G$ is calculated by the Pythagoras´ theorem (eq. S2, F indicates the different dyes) with the distance between two oligonucleotides o (0.34 nm), the diameter of a helix d (2 nm) and the crossover between two helix c (1 nm).

$$d_F = \sqrt{((x_S - x_{P_F}) \cdot d + 3c\,)^2 + ((y_S - y_{P_F}) \cdot o)^2} \quad (S2)$$

The distances are 10.98 nm for S-$P_R$ ($d_R$) and 9.79 nm for S-$P_G$ ($d_G$). The height difference, h, is the sum of linker between dye and DNA origami structure (0.5 nm), the diameter of the DNA origami structure (2 nm), the crossover between DNA origami structure and formed linking helix (1 nm), the diameter of the linking helix (2 nm) and linker between linking helix and NP (0.5 nm), so overall 6 nm. By using the Pythagoras´ theorem a second time and subtract the radius r of the NP, a is calculated by Equation (S3).

$$a_{F,r} = \sqrt{((h\, + r)^2 + \,{d_F}^2)} - r \qquad (S3)$$

The overall distances are shown in table S2.

Table S2: Distances calulations between NP surfaces and both dyes (Atto647N and Atto532).

| r [nm] | $a_{G,r}$ [nm] | $A_{R,r}$ [nm] |
|---|---|---|
| 5 | 10.8 | 11.7 |
| 10 | 10.0 | 10.8 |
| 15 | 9.4 | 10.1 |
| 20 | 9.0 | 9.6 |

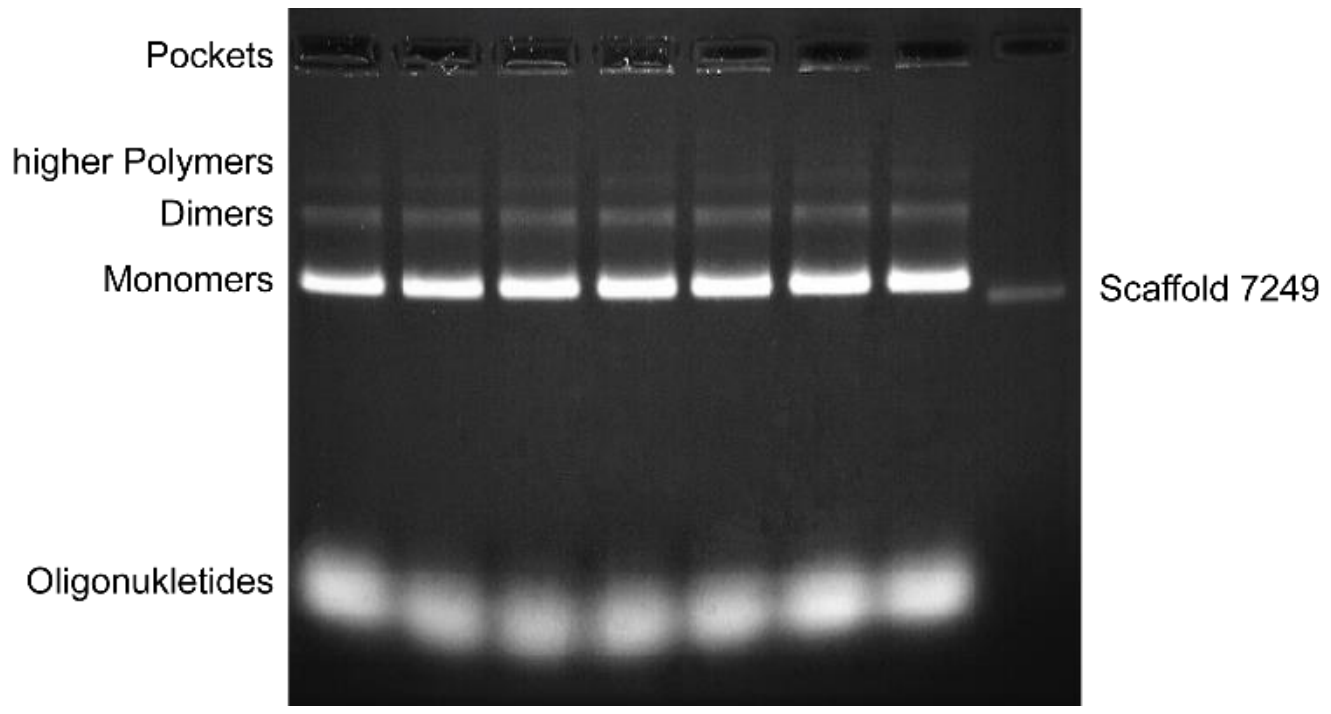

Figure S4: Gel images for the purified rectangular DNA origami structures with monomers, polymers, oligonucleotide and the scaffold as a reference.

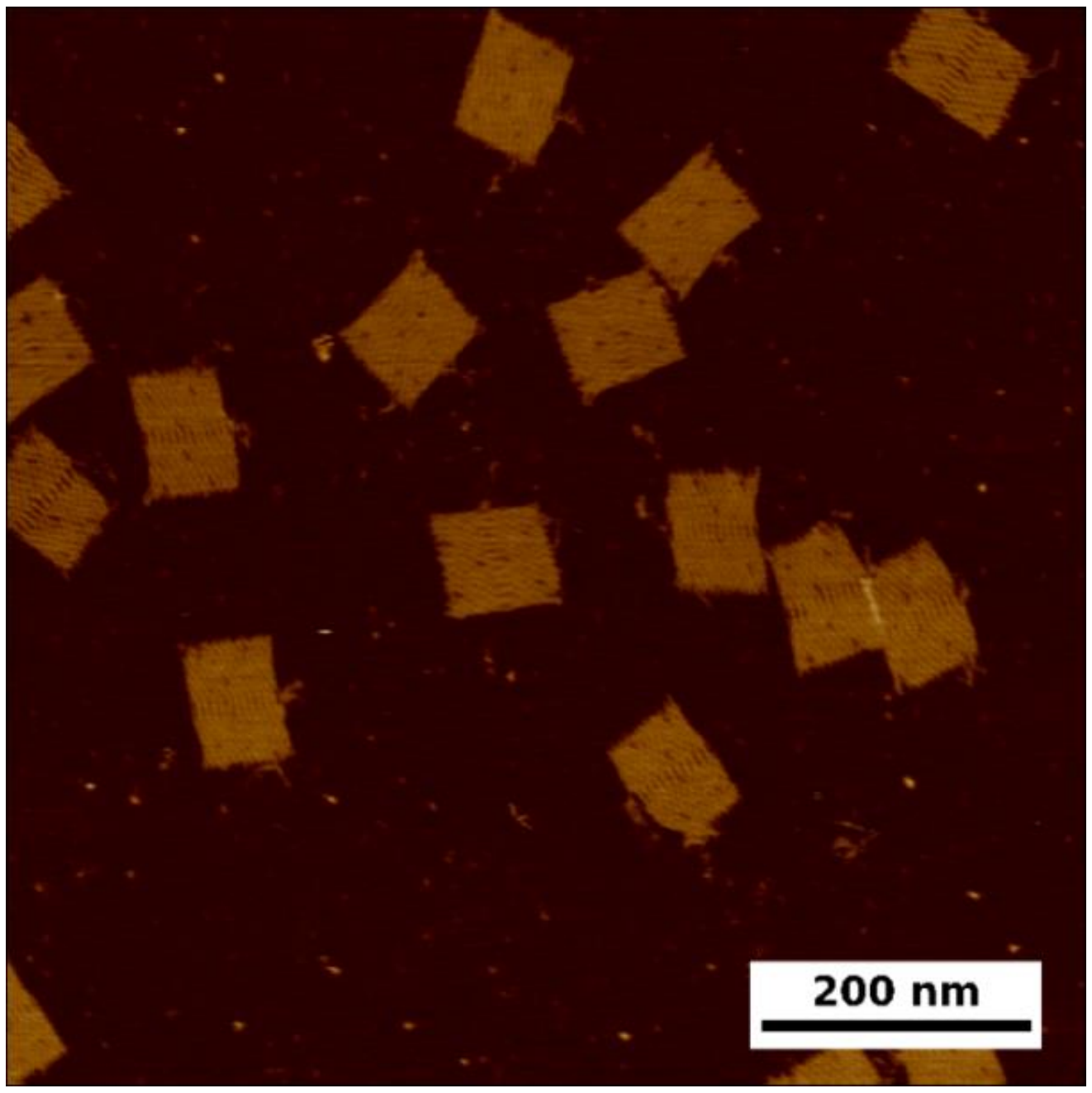

Figure S5: 800×800 µm images of the rectangular DNA origami structure. The holes in the edges and on the left and right side from sprout like center are showing the eight missing oligonucleotides from biotin.

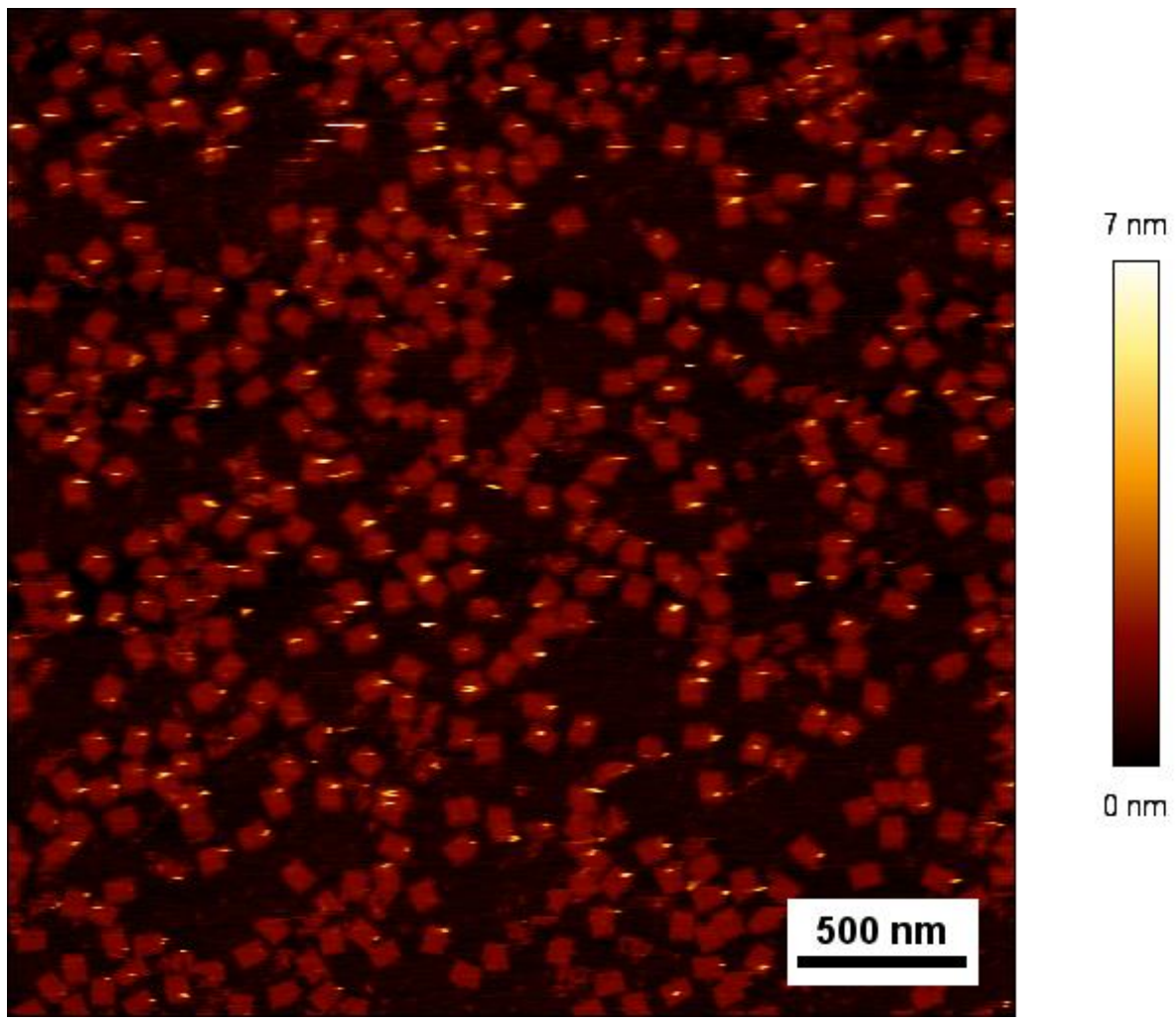

**Figure S6:** Rectangular DNA origami structure with 5 nm gold nanoparticle with a scale bar ranging from 0 to 7 nm. This DNA origami structure has an height with NP of 2 nm (one helix).

Table S3: Volume of the oligonucleotides with a thiol group at the 3` for nanoparticle with different sizes and materials.

| d [nm] | 5 Au | 10 Au | 15 Au | 20 Au |
|---|---|---|---|---|
| V [µL/mL] | 95.4 | 49.5 | 31.7 | 24 |

Table S4: Salting steps.

| Step | 1 | 2 | 3 | 4 | 5 | 6 | 7 |
|---|---|---|---|---|---|---|---|
| V [µL] | 10 | 10 | 20 | 20 | 20 | 20 | 50 |
| Step | 8 | 9 | 10 | 11 | 12 | 13 | |
| V [µL] | 50 | 50 | 50 | 100 | 100 | 100 | |

## 4. Design of DNA origami structure

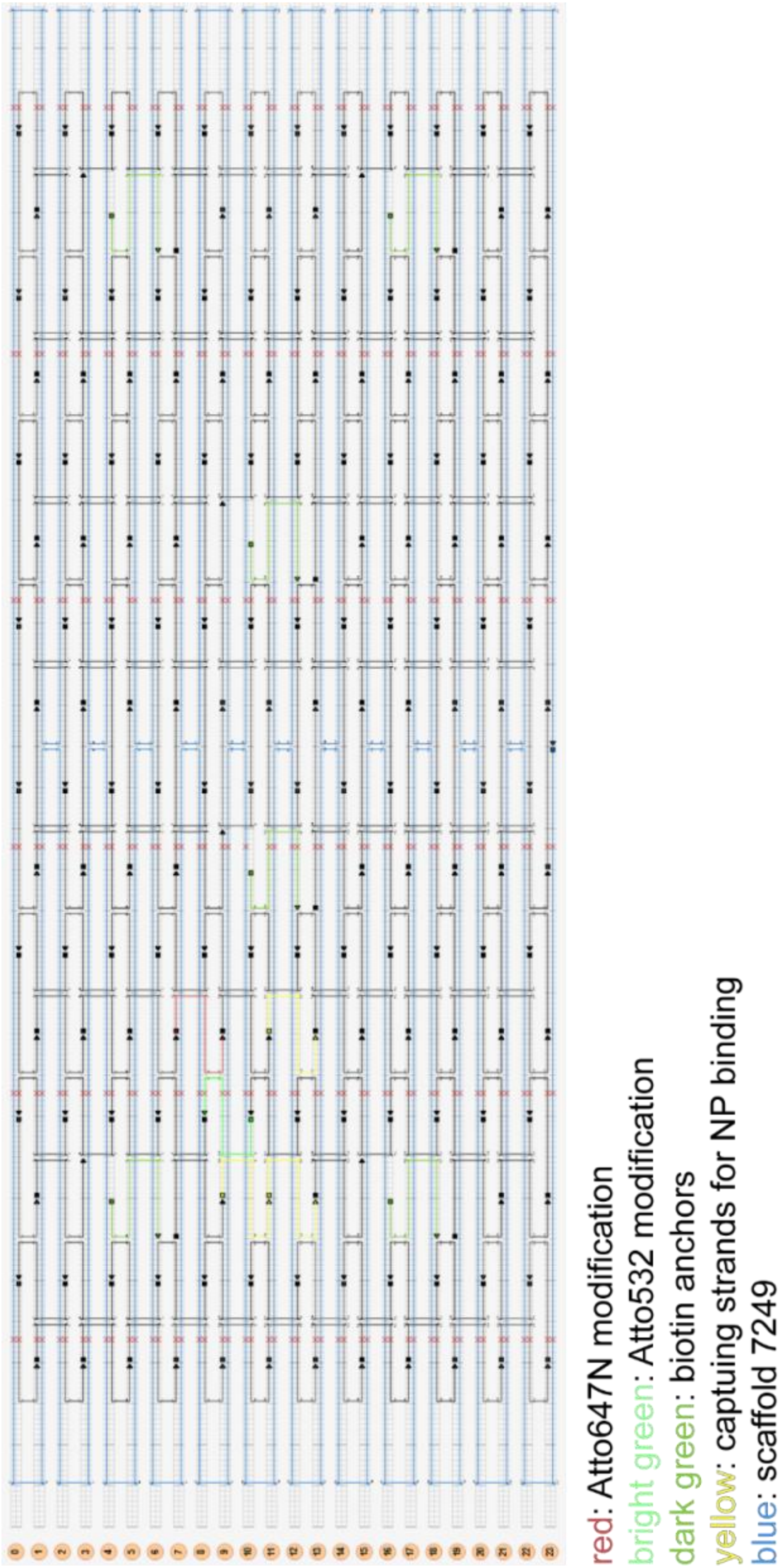

Figure S7: caDNAno image of rectangular DNA origami structure.

Tab. S 5. sequences of unmodified staples.

| Sequence (5‘->3‘) | Length [nt] |
|---|---|
| TGACAACTCGCTGAGGCTTGCATTATACCA | 30 |
| AGAAAACAAAGAAGATGATGAAACAGGCTGCG | 32 |
| CTGTAGCTTGACTATTATAGTCAGTTCATTGA | 32 |
| TATATTTTGTCATTGCCTGAGAGTGGAAGATTGTATAAGC | 40 |
| CTTTAGGGCCTGCAACAGTGCCAATACGTG | 30 |
| TTAATGAACTAGAGGATCCCCGGGGGGTAACG | 32 |
| TCATCGCCAACAAAGTACAACGGACGCCAGCA | 32 |
| TCTTCGCTGCACCGCTTCTGGTGCGGCCTTCC | 32 |
| CTACCATAGTTTGAGTAACATTTAAAATAT | 30 |
| CGAAAGACTTTGATAAGAGGTCATATTTCGCA | 32 |
| ATTTTAAAATCAAAATTATTTGCACGGATTCG | 32 |
| GCGAAAAATCCCTTATAAATCAAGCCGGCG | 30 |
| CTGTGTGATTGCGTTGCGCTCACTAGAGTTGC | 32 |
| AGCGCGATGATAAATTGTGTCGTGACGAGA | 30 |
| GATGGTTTGAACGAGTAGTAAATTTACCATTA | 32 |
| GATGTGCTTCAGGAAGATCGCACAATGTGA | 30 |
| TAAATCAAAATAATTCGCGTCTCGGAAACC | 30 |
| GACAAAAGGTAAAGTAATCGCCATATTTAACAAAACTTTT | 40 |
| CCAGGGTTGCCAGTTTGAGGGGACCCGTGGGA | 32 |
| CTTATCATTCCCGACTTGCGGGAGCCTAATTT | 32 |
| CAGAAGATTAGATAATACATTTGTCGACAA | 30 |
| CGTAAAACAGAAATAAAAATCCTTTGCCCGAAAGATTAGA | 40 |
| AATACTGCCCAAAAGGAATTACGTGGCTCA | 30 |
| ATATTCGGAACCATCGCCCACGCAGAGAAGGA | 32 |
| ATACATACCGAGGAAACGCAATAAGAAGCGCATTAGACGG | 40 |
| CATCAAGTAAAACGAACTAACGAGTTGAGA | 30 |
| TTTCGGAAGTGCCGTCGAGAGGGTGAGTTTCG | 32 |
| AATAGTAAACACTATCATAACCCTCATTGTGA | 32 |
| GACCTGCTCTTTGACCCCCAGCGAGGGAGTTA | 32 |
| AACACCAAATTTCAACTTTAATCGTTTACC | 30 |
| CTCGTATTAGAAATTGCGTAGATACAGTAC | 30 |
| ATTACCTTTGAATAAGGCTTGCCCAAATCCGC | 32 |
| GCCGTCAAAAAACAGAGGTGAGGCCTATTAGT | 32 |
| AGTATAAAGTTCAGCTAATGCAGATGTCTTTC | 32 |
| TGTAGCCATTAAAATTCGCATTAAATGCCGGA | 32 |
| CAGCGAAACTTGCTTTCGAGGTGTTGCTAA | 30 |
| TACCGAGCTCGAATTCGGGAAACCTGTCGTGCAGCTGATT | 40 |
| GCGGATAACCTATTATTCTGAAACAGACGATT | 32 |
| AGCAAGCGTAGGGTTGAGTGTTGTAGGGAGCC | 32 |
| TTAAAGCCAGAGCCGCCACCCTCGACAGAA | 30 |
| TTCCAGTCGTAATCATGGTCATAAAAGGGG | 30 |
| CACAACAGGTGCCTAATGAGTGCCCAGCAG | 30 |
| TCAAGTTTCATTAAAGGTGAATATAAAAGA | 30 |
| GCTTTCCGATTACGCCAGCTGGCGGCTGTTTC | 32 |
| CCACCCTCTATTCACAAACAAATACCTGCCTA | 32 |
| TCAAATATAACCTCCGGCTTAGGTAACAATTT | 32 |
| AAAGGCCGGAGACAGCTAGCTGATAAATTAATTTTTGT | 38 |
| CTGAGCAAAAATTAATTACATTTTGGGTTA | 30 |
| GCGGAACATCTGAATAATGGAAGGTACAAAAT | 32 |
| CACCAGAAAGGTTGAGGCAGGTCATGAAAG | 30 |
| GAAATTATTGCCTTTAGCGTCAGACCGGAACC | 32 |
| GAATTTATTTAATGGTTTGAAATATTCTTACC | 32 |
| GTACCGCAATTCTAAGAACGCGAGTATTATTT | 32 |
| GTTTATCAATATGCGTTATACAAACCGACCGTGTGATAAA | 40 |
| CAACTGTTGCGCCATTCGCCATTCAAACATCA | 32 |
| AAAGTCACAAAATAAACAGCCAGCGTTTTA | 30 |
| CAGGAGGTGGGGTCAGTGCCTTGAGTCTCTGAATTTACCG | 40 |
| GTAATAAGTTAGGCAGAGGCATTTATGATATT | 32 |
| ATTATACTAAGAAACCACCAGAAGTCAACAGT | 32 |
| GAGGGTAGGATTCAAAAGGGTGAGACATCCAA | 32 |
| AAGGAAACATAAAGGTGGCAACATTATCACCG | 32 |
| TTTTATTTAAGCAAATCAGATATTTTTTGT | 30 |
| TAGGTAAACTATTTTTGAGAGATCAAACGTTA | 32 |
| ACAAACGGAAAAGCCCCAAAAACACTGGAGCA | 32 |
| ATACCCAACAGTATGTTAGCAAATTAGAGC | 30 |
| ACCGATTGTCGGCATTTTCGGTCATAATCA | 30 |
| CATAAATCTTTGAATACCAAGTGTTAGAAC | 30 |
| TATAACTAACAAAGAACGCGAGAACGCCAA | 30 |
| ACGGCTACAAAAGGAGCCTTTAATGTGAGAAT | 32 |
| TTAGGATTGGCTGAGACTCCTCAATAACCGAT | 32 |
| AATTGAGAATTCTGTCCAGACGACTAAACCAA | 32 |
| AATAGCTATCAATAGAAAATTCAACATTCA | 30 |
| ACCTTGCTTGGTCAGTTGGCAAAGAGCGGA | 30 |
| ATATTTTGGCTTTCATCAACATTATCCAGCCA | 32 |
| AGGCTCCAGAGGCTTTGAGGACACGGGTAA | 30 |
| GCAAGGCCTCACCAGTAGCACCATGGGCTTGA | 32 |
| TTAACACCAGCACTAACAACTAATCGTTATTA | 32 |
| GCCAGTTAGAGGGTAATTGAGCGCTTTAAGAA | 32 |
| TTTATCAGGACAGCATCGGAACGACACCAACCTAAAACGA | 40 |
| TTGACAGGCCACCACCAGAGCCGCGATTTGTA | 32 |
| AGACGACAAAGAAGTTTTGCCATAATTCGAGCTTCAA | 37 |
| CGATAGCATTGAGCCATTTGGGAACGTAGAAA | 32 |
| ACACTCATCCATGTTACTTAGCCGAAAGCTGC | 32 |
| TGGAACAACCGCCTGGCCCTGAGGCCCGCT | 30 |
| TTATACCACCAAATCAACGTAACGAACGAG | 30 |
| TAATCAGCGGATTGACCGTAATCGTAACCG | 30 |
| CGCGCAGATTACCTTTTTTAATGGGAGAGACT | 32 |
| GTTTATTTTGTCACAATCTTACCGAAGCCCTTTAATATCA | 40 |
| AAATCACCTTCCAGTAAGCGTCAGTAATAA | 30 |
| TGAAAGGAGCAAATGAAAAATCTAGAGATAGA | 32 |
| CCTGATTGCAATATATGTGAGTGATCAATAGT | 32 |
| CTTAGATTTAAGGCGTTAAATAAAGCCTGT | 30 |
| AAGTAAGCAGACACCACGGAATAATATTGACG | 32 |
| TTATTACGAAGAACTGGCATGATTGCGAGAGG | 32 |
| GGCCTTGAAGAGCCACCACCCTCAGAAACCAT | 32 |
| GCCATCAAGCTCATTTTTTAACCACAAATCCA | 32 |
| TTGCTCCTTTCAAATATCGCGTTTGAGGGGGT | 32 |
| TTAACGTCTAACATAAAAACAGGTAACGGA | 30 |
| AGGCAAAGGGAAGGGCGATCGGCAATTCCA | 30 |
| ATCCCAATGAGAATTAACTGAACAGTTACCAG | 32 |
| AAAGCACTAAATCGGAACCCTAATCCAGTT | 30 |
| ATCCCCCTATACCACATTCAACTAGAAAAATC | 32 |
| TCATTCAGATGCGATTTTAAGAACAGGCATAG | 32 |
| GCGAACCTCCAAGAACGGGTATGACAATAA | 30 |
| TAAATGAATTTTCTGTATGGGATTAATTTCTT | 32 |
| TCACCGACGCACCGTAATCAGTAGCAGAACCG | 32 |
| CATTTGAAGGCGAATTATTCATTTTTGTTTGG | 32 |
| ACAACATGCCAACGCTCAACAGTCTTCTGA | 30 |
| TCACCAGTACAAACTACAACGCCTAGTACCAG | 32 |
| GCCCGAGAGTCCACGCTGGTTTGCAGCTAACT | 32 |
| GCGCAGACAAGAGGCAAAAGAATCCCTCAG | 30 |
| ATTATCATTCAATATAATCCTGACAATTAC | 30 |
| AAACAGCTTTTTGCGGGATCGTCAACACTAAA | 32 |
| ACCCTTCTGACCTGAAAGCGTAAGACGCTGAG | 32 |
| GTATAGCAAACAGTTAATGCCCAATCCTCA | 30 |
| AAGGCCGCTGATACCGATAGTTGCGACGTTAG | 32 |
| CCTAAATCAAAATCATAGGTCTAAACAGTA | 30 |
| CTTTTGCAGATAAAAACCAAAATAAAGACTCC | 32 |
| CTTTTACAAAATCGTCGCTATTAGCGATAG | 30 |
| CATGTAATAGAATATAAAGTACCAAGCCGT | 30 |
| GACCAACTAATGCCACTACGAAGGGGGTAGCA | 32 |
| CAGCAAAAGGAAACGTCACCAATGAGCCGC | 30 |
| TAAATCGGGATTCCCAATTCTGCGATATAATG | 32 |
| AACGCAAAGATAGCCGAACAAACCCTGAAC | 30 |
| TAAATCATATAACCTGTTTAGCTAACCTTTAA | 32 |
| ATCGCAAGTATGTAAATGCTGATGATAGGAAC | 32 |
| AGCCAGCAATTGAGGAAGGTTATCATCATTTT | 32 |
| GCCCTTCAGAGTCCACTATTAAAGGGTGCCGT | 32 |
| GCTATCAGAAATGCAATGCCTGAATTAGCA | 30 |
| GCGAGTAAAAATATTTAAATTGTTACAAAG | 30 |
| TATTAAGAAGCGGGGTTTTGCTCGTAGCAT | 30 |
| AATACGTTTGAAAGAGGACAGACTGACCTT | 30 |
| AAATTAAGTTGACCATTAGATACTTTTGCG | 30 |
| TGCATCTTTCCCAGTCACGACGGCCTGCAG | 30 |
| TACGTTAAAGTAATCTTGACAAGAACCGAACT | 32 |
| ATGCAGATACATAACGGGAATCGTCATAAATAAAGCAAAG | 40 |
| CCCGATTTAGAGCTTGACGGGGAAAAAGAATA | 32 |
| ACCTTTTTATTTTAGTTAATTTCATAGGGCTT | 32 |
| CACATTAAAATTGTTATCCGCTCATGCGGGCC | 32 |
| GCCTCCCTCAGAATGGAAAGCGCAGTAACAGT | 32 |
| ACAACTTTCAACAGTTTCAGCGGATGTATCGG | 32 |
| CTTTAATGCGCGAACTGATAGCCCCACCAG | 30 |
| GCACAGACAATATTTTTGAATGGGGTCAGTA | 31 |
| AGAAAGGAACAACTAAAGGAATTCAAAAAAA | 31 |
| AACAGTTTTGTACCAAAAACATTTTATTTC | 30 |
| AGGAACCCATGTACCGTAACACTTGATATAA | 31 |
| CCAACAGGAGCGAACCAGACCGGAGCCTTTAC | 32 |
| AACGCAAAATCGATGAACGGTACCGGTTGA | 30 |
| CAACCGTTTCAAATCACCATCAATTCGAGCCA | 32 |
| TTCTACTACGCGAGCTGAAAAGGTTACCGCGC | 32 |
| GCCTTAAACCAATCAATAATCGGCACGCGCCT | 32 |
| GCCCGTATCCGGAATAGGTGTATCAGCCCAAT | 32 |
| TCCACAGACAGCCCTCATAGTTAGCGTAACGA | 32 |
| TCTAAAGTTTTGTCGTCTTTCCAGCCGACAA | 31 |
| AACAAGAGGGATAAAAATTTTTAGCATAAAGC | 32 |
| AGAGAGAAAAAAATGAAAATAGCAAGCAAACT | 32 |
| TCAATATCGAACCTCAAATATCAATTCCGAAA | 32 |
| CCACCCTCATTTTCAGGGATAGCAACCGTACT | 32 |
| GTCGACTTCGGCCAACGCGCGGGGTTTTTC | 30 |
| GTTTTAACTTAGTACCGCCACCCAGAGCCA | 30 |
| TTAGTATCACAATAGATAAGTCCACGAGCA | 30 |
| GCAATTCACATATTCCTGATTATCAAAGTGTA | 32 |
| TAAAAGGGACATTCTGGCCAACAAAGCATC | 30 |
| AAGCCTGGTACGAGCCGGAAGCATAGATGATG | 32 |
| AACGTGGCGAGAAAGGAAGGGAAACCAGTAA | 31 |
| CCAATAGCTCATCGTAGGAATCATGGCATCAA | 32 |
| ACGCTAACACCCACAAGAATTGAAAATAGC | 30 |
| TGTAGAAATCAAGATTAGTTGCTCTTACCA | 30 |
| CAAATCAAGTTTTTTGGGGTCGAAACGTGGA | 31 |
| TCGGCAAATCCTGTTTGATGGTGGACCCTCAA | 32 |
| TTTTCACTCAAAGGGCGAAAAACCATCACC | 30 |
| CTCCAACGCAGTGAGACGGGCAACCAGCTGCA | 32 |
| TTTACCCCAACATGTTTTAAATTTCCATAT | 30 |
| GAGAGATAGAGCGTCTTTCCAGAGGTTTTGAA | 32 |
| TTTAGGACAAATGCTTTAAACAATCAGGTC | 30 |

Tab. S 6: Modified staples with dyes, biotin and capturing strands for NP.

| Sequence (5‘->3‘) | Length [nt] |
|---|---|
| TAAGAGCAAATGTTTAGACTGGATAG-Atto647N-AAGCC | 32 |
| GATGGCTTATCAAAA-Atto532-GATTAAGAGCGTCC | 30 |
| Biotin-CGGATTCTGACGACAGTATCGGCCGCAAGGCGATTAAGTT | 40 |
| Biotin-AGCCACCACTGTAGCGCGTTTTCAAGGGAGGGAAGGTAAA | 40 |
| Biotin-ATAAGGGAACCGGATATTCATTACGTCAGGACGTTGGGAA | 40 |
| Biotin-GAGAAGAGATAACCTTGCTTCTGTTCGGGAGAAACAATAA | 40 |
| Biotin-TAGAGAGTTATTTTCATTTGGGGATAGTAGTAGCATTA | 38 |
| Biotin-GAAACGATAGAAGGCTTATCCGGTCTCATCGAGAACAAGC | 40 |
| AATGGTCAACAGGCAAGGCAAAGAGTAATGTGAAAAAAAAAAAAAAAAAAAA | 52 |
| GATTTAGTCAATAAAGCCTCAGAGAACCCTCAAAAAAAAAAAAAAAAAAAAA | 52 |
| CGGATTGCAGAGCTTAATTGCTGAAACGAGTAAAAAAAAAAAAAAAAAAAAA | 52 |

Oligonucleotide sequence for nanoparticle from 5' to 3':
TTTTTTTTTTTTTTTTTTTTTTTTT-Thiol

# 5. Numerical calculations

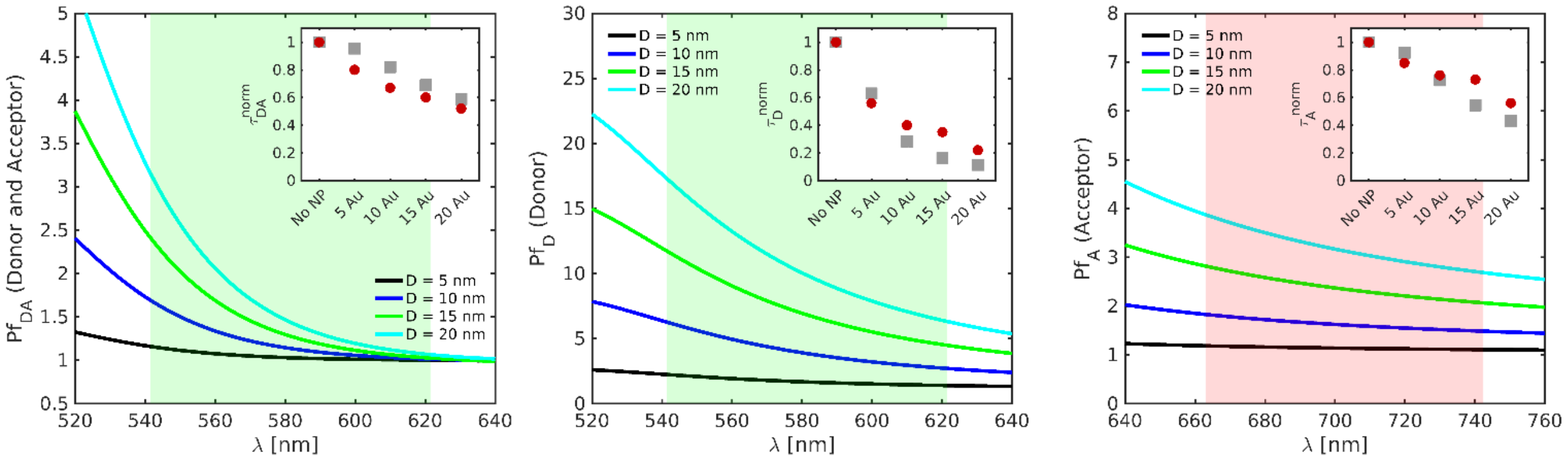

Figure S8: Numerical Purcell factor, $Pf$, spectra for the donor in presence (left) and absence (center) of the acceptor, and for the acceptor in isolation (right). Calculations for the four Au NP sizes considered in the experiments are shown ($D$ indicates the NP diameter). The insets show normalized lifetimes calculated from the spectral averaging (taken within the colored range in the main panels) of the $Pf$ spectra and using Equation (4). Experimental and theoretical results are plotted in red circles and grey squares, respectively.

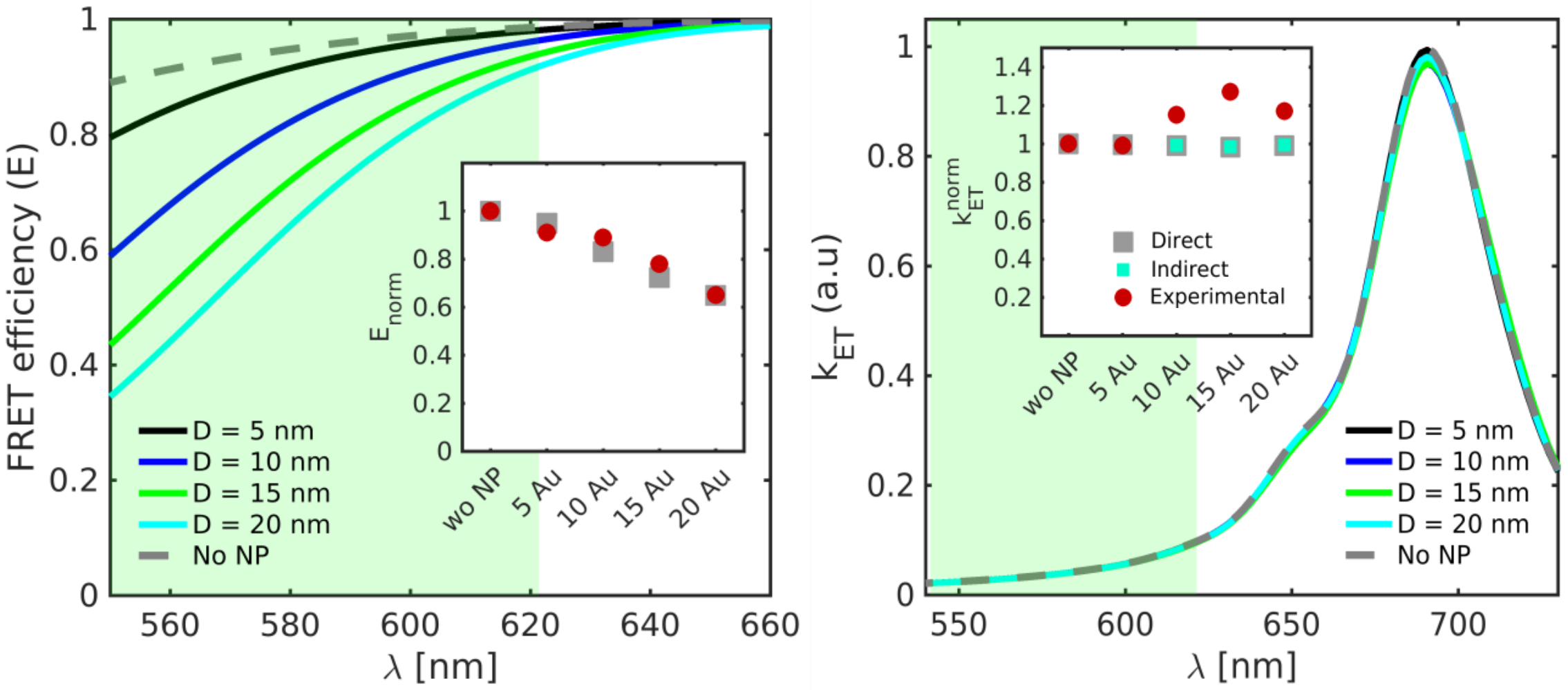

Figure S9: Theoretical predictions for the FRET efficiency and rate. Right: $E = 1 - Pf_D/Pf_{DA}$ (note the equivalence with Equation (3)) as a function of the donor emission wavelength. The inset (grey squares) plots the efficiency obtained from the spectral averaging within the green window. Left: $k_{ET} \propto V^{-1} \int |E_{DA}|^2 dV$ as a function of the donor emission wavelength. The inset (grey squares) shows the rate obtained from the spectral averaging within the green window. For comparison, the indirect prediction obtained from the evaluation of Equation (2) with numerical results in the insets of Figure S8 is shown in cyan squares. In both panels, red circles correspond to experimental data.